\documentclass[preprint,aps,amsmath,amssymb]{revtex4}

\newcommand {\bc}{\begin{center}}
\newcommand {\ec}{\end{center}}
\newcommand {\bea}{\begin{eqnarray}}
\newcommand {\eea}{\end{eqnarray}}
\newcommand {\be}{\begin{equation}}
\newcommand {\ee}{\end{equation}}

\def\lsim{\mathrel{\rlap{\lower4pt\hbox{\hskip1pt$\sim$}}
    \raise1pt\hbox{$<$}}}               
\def\gsim{\mathrel{\rlap{\lower4pt\hbox{\hskip1pt$\sim$}}
    \raise1pt\hbox{$>$}}}

\usepackage[mathscr]{euscript}
\usepackage{graphicx}

\begin{document}


\title{Second order fluid dynamics for the unitary Fermi gas
from kinetic theory}

\author{Thomas Sch\"afer}

\affiliation{Department of Physics, North Carolina State University,
Raleigh, NC 27695}

\begin{abstract}
We compute second order transport coefficients of the dilute 
Fermi gas at unitarity. The calculation is based on kinetic theory 
and the Boltzmann equation at second order in the Knudsen expansion.
The second order transport coefficients describe the shear stress
relaxation time, non-linear terms in the strain-stress relation,
and non-linear couplings between vorticity and strain. An exact
calculation in the dilute limit gives $\tau_R=\eta/P$, where $\tau_R$
is the shear stress relaxation time, $\eta$ is the shear viscosity,
and $P$ is pressure. This relation is identical to the result 
obtained using the Bhatnagar-Gross-Krook (BGK) approximation to
the collision term, but other transport coefficients are sensitive
to the exact collision integral.  
\end{abstract}

\maketitle

\section{Introduction}
\label{sec_intro}
 
 The dilute Fermi gas at unitarity provides a new paradigm for
transport properties of strongly correlated quantum fluids 
\cite{Schafer:2009dj,Adams:2012th,Schaefer:2014awa}.  
Over the last several years there have been several experimental 
\cite{oHara:2002,Kinast:2004b,Bartenstein:2004,Schafer:2007pr,Cao:2010wa,Elliott:2013,Elliott:2013b}
and theoretical \cite{Massignan:2004,Gelman:2004fj,Bruun:2005,Bruun:2006,Bruun:2007,Rupak:2007vp,Son:2008ye,Herzog:2008wg,Enss:2010qh,Guo:2010,Hofmann:2011qs,Enss:2012,Levin:2013}
studies devoted to the shear viscosity of the unitary Fermi gas. It 
was found that the system behaves as a nearly perfect fluid, which 
means that the ratio of entropy density to shear viscosity is close 
to the quantum bound $\eta/s=\hbar/(4\pi k_B)$ \cite{Kovtun:2004de}.
This bound was derived using the holographic duality between string
theory in ten space-time dimensions, and field theory in four or fewer 
dimensions. 

 The main difficulty in providing more accurate determinations of 
the temperature and density dependence of $\eta/s$ is that experiments
mainly constrain the average value of the viscosity of a harmonically
trapped gas cloud. For the range of temperatures probed in experiments
the center of the cloud behaves hydrodynamically, but in the dilute 
corona the mean free path is large and hydrodynamics is not applicable,
see Sect.~\ref{sec_scales}. The theoretical challenge is to describe 
the transition between the hydrodynamic regime and the ballistic corona. 
A possible approach to this problem is to take into account a non-zero 
dissipative relaxation time \cite{Schaefer:2009px,Bruun:2007}. The 
relaxation time describes the time scale for dissipative stresses to 
approach the value given by the Navier-Stokes theory. In the dilute
corona the relaxation time $\tau_R\sim l_{\it mfp}$ is large. As a 
consequence dissipative stresses remain small, and the rate of 
dissipation smoothly approaches the ballistic limit. 

 In a systematic approach to fluid dynamics the relaxation time 
appears at second in the gradient expansion of the stress tensor
\cite{Chao:2011cy}. The corresponding kinetic coefficients can 
be determined using kinetic theory at second order in the 
Knudsen expansion. A simple estimate for the viscous relaxation 
time can be obtained using the the Bhatnagar-Gross-Krook (BGK)
approximation to the collision integral \cite{Bhatnagar:1954}. In 
this approximation, all departures from equilibrium are assumed to 
relax at the same rate $1/\tau_0$. Not surprisingly, one finds that 
the viscous relaxation time $\tau_R$ is given by $\tau_R=\tau_0$.
Using the result for the shear viscosity in the BGK theory, $\eta
=P\tau_0$, we can write $\tau_R=\eta/P$. The ideal gas  law $P\sim 
nT$ together with the fact that the viscosity of a dilute gas is 
independent of density implies that the relaxation time is inversely 
proportional to density. 

 The BGK estimate $\tau_R=\eta/P$ has been rederived many times
in the literature, see  \cite{Bruun:2007,Braby:2010tk,Chao:2011cy},
but it is not known how reliable this approximation is. Our goal in 
this paper is to provide an exact calculation of $\tau_R$ in the 
dilute limit. The calculation is based on kinetic theory and the 
Boltzmann equation at leading order in the fugacity of the gas. We 
also determine other second order transport coefficient related to 
the shear stress, in particular the leading non-linear terms in 
strain-stress relation, and the mixing between strain and vorticity. 
For simplicity we do not compute second order transport coefficients 
related to heat flow. In the cold atom experiments the cloud is 
initially isothermal, and temperature gradients are only generated 
by viscous heating. As a result, the effect of thermal conductivity 
on the motion of the gas is already a second order effect, and higher 
order corrections are expected to be very small. 

\section{Scales and expansion parameters}
\label{sec_scales}
 
 Dilute atomic gases are characterized by the condition $k_Fr\ll 1$,
where $r$ is the range of the interaction, and $k_F$ is the Fermi 
momentum. The Fermi momentum is defined in terms of the density, $n=
\nu k_F^3/(6\pi^2)$, where $\nu$ is the number of species. We are mostly 
interested in the case $\nu=2$, which describes an unpolarized
two-component gas. The unitary Fermi gas corresponds to the limit $k_Fa\to\
\infty$, where $a$ is the $s$-wave scattering length. This implies that 
the gas is dilute, but very strongly correlated.  
 
 The density of the gas determines a temperature scale, $k_B T_F=k_F^2/
(2m)$ \footnote{Note that in the following we will set the Boltzmann 
constant $k_B$ and Planck's constant $\hbar$ equal to one.}. For $T<T_F$ 
the gas is a strongly correlated quantum fluid, and the only reliable 
theoretical approach to equilibrium and non-equilibrium properties of 
the unitary Fermi gas is the quantum Monte Carlo (QMC) method. In the 
case of thermodynamic properties, QMC computations are very successful 
\cite{VanHoucke:2011ux}, but calculations of non-equilibrium properties 
are more challenging \cite{Wlazlowski:2012jb,Wlazlowski:2013owa}. The 
unitary gas has a second order phase transition to a superfluid phase. 
The most precise determination of the critical temperature comes from 
experiments with trapped atomic gases, which give $T_c/T_F = 0.167(13)$ 
\cite{Ku:2011}.

 At high temperature, $T>T_F$, the thermal de Broglie wave length 
$\lambda_{\it dB}=[2\pi/(mT)]^{1/2}$ of the particles is small, and the 
quantum diluteness of the gas $n\lambda_{\it dB}^3$ can be used as an 
expansion parameter. In the case of thermodynamic properties, this is 
the familiar virial expansion. Note that the fugacity of the gas is 
given by $z=e^{\beta\mu}\simeq (n\lambda_{\it dB}^3)/\nu$, and the virial 
expansion is equivalent to an expansion in powers of $z$. At leading
order the equation of state is that of an ideal gas, $P=nT$, and the 
first non-trivial correction is governed by the second virial 
coefficient $b_2$. In the weak coupling limit $b_2\sim a/\lambda$,
but in the $a\to\infty$ limit $b_2$ is a pure number. Experimental 
data show that at unitarity the viral expansion describes the equation
of state for $z\lsim 1$ \cite{Nascimbene:2009}.

 In the high temperature limit transport properties of the unitary
gas can be understood in terms of kinetic theory and the Boltzmann
equation. Kinetic theory is based on the existence of well-defined 
quasi-particles, and requires that the quasi-particle width $\Gamma$
is small compared to the mean quasi-particle energy. In the high 
temperature limit $\Gamma\sim zT \ll E \sim T$ \cite{Schaefer:2013oba}.
There are no complete, rigorous, calculations beyond leading order 
in $z$ in the literature, and as result the regime of validity of 
kinetic theory is difficult to establish. Experiments and QMC 
calculations are consistent with the idea that the range of 
convergence is similar to that of the virial expansion, $z\lsim 1$ 
\cite{Cao:2010wa,Wlazlowski:2013owa}. 

 Experiments involve a larger number of atoms, typically on the 
order of a few times $10^6$. This implies that the quasi-particle 
distribution function varies smoothly over the size $L$ of the trap. 
The microscopic scale is given by the mean free path $l_{\it mfp}$, and 
the expansion parameter is the Knudsen number ${\it Kn}=l_{\it mfp}/L$. 
The mean free path is given by $l_{\it mfp}=1/(n\sigma)$, where $n$ is 
the density and $\sigma$ is the two-body cross section. In the unitary 
gas $\sigma=4\pi/q^2$, where $q$ is the momentum transfer. A simple 
estimate of the Knudsen number can be obtained by averaging $\sigma$ 
over a thermal distribution. The Knudsen number at the center of the trap 
is \cite{Adams:2012th}
\be
\label{Kn_0}
{\it Kn}(0) = \frac{3\pi^{1/2}}{4(3\lambda N)^{1/3}}
 \left(\frac{T}{T^{\it trap}_F}\right)^2\, ,
\ee
where $\lambda$ is the aspect ratio of the trap, and $T^{\it trap}_F$ is 
the Fermi temperature of a non-interacting Fermi gas at the center of 
the trap. For the parameters probed in experiments ${\it Kn}(0)\ll 1$ 
corresponds to $T\lsim 5T_F$. Since the mean cross section is only 
a function of temperature the mean free path in the trap scales as 
$l_{\it mfp}\sim n^{-1}$. This implies that 
\be 
\label{Kn_x}
{\it Kn}(x) \simeq {\it Kn}(0)
   \exp\left( \frac{m}{2T}\sum_i \omega_i^2 x_i^2\right)\, , 
\ee
where $\omega_i$ ($i=1,2,3$) are the trapping frequencies. Near the 
edge of the trap the Knudsen number becomes large, and the expansion
in gradients of the distribution function breaks down. Equ.~(\ref{Kn_x})
implies that the relevant scale is close to mean square cloud radius.

 The estimates in equ.~(\ref{Kn_0},\ref{Kn_x}) refer to a static 
trapped Fermi gas. Conformal invariance implies that, up to 
dissipative effects, an expanding gas cloud evolves by a time 
dependent scale transformation. This means that in a co-moving 
frame dimensionless ratios such as the Knudsen number are independent
of time. In particular, if ${\it Kn}(0)\ll 1$ initially, then the
Knudsen expansion remains valid throughout the evolution of the 
cloud. 

 In the regime ${\it Kn}\ll 1$ kinetic theory is equivalent to 
fluid dynamics. Fluid dynamics is based on the assumption of local
thermodynamic equilibrium, and on the fact that thermodynamic 
variables vary smoothly over the extent of the system. Gradients of
thermodynamic variables determine dissipative corrections to the 
equations of ideal fluid dynamics. The expansion parameter that 
controls the validity of fluid dynamics is the ratio of dissipative
to ideal contributions in the energy and momentum currents. The main 
parameter is the inverse Reynolds number
\be 
 {\it Re}^{-1}=\frac{\eta}{\rho L u}\, , 
\ee
which measures the ratio of dissipative and ideal contributions
to the stress tensor. Here, $\eta$ is the shear viscosity, $\rho$ 
is the mass density, and $u$ is the velocity of the fluid. In the 
kinetic regime ${\it Kn}\simeq {\it Re}^{-1}$, and the gradient 
expansion in kinetic theory is equivalent to the expansion in 
gradients of thermodynamic variables \cite{Schaefer:2014awa}. The 
fluid dynamical description is valuable because the gradient 
expansion remains valid even in the regime $T\lsim T_F$ where 
the quasi-particle width is comparable to the quasi-particle 
energy and kinetic theory is not applicable.

 In the case of the experiments described in  
\cite{oHara:2002,Elliott:2013,Elliott:2013b} fluid dynamics is 
applicable at the center of the trap, but kinetic theory is not. 
The mean free path is comparable to the inter-particle spacing, 
and $\eta/s \ll 1$. Fluid dynamics breaks down in the dilute part 
of the cloud, but in this regime kinetic theory is reliable. In order 
to study the transition between fluid dynamics and kinetic theory 
we will compute the leading second order gradient corrections 
to the Navier-Stokes equation. These terms can used to quantify 
the breakdown of the Navier-Stokes theory. In Sect.~\ref{sec_out} 
we will show that one can resum second order gradient corrections,
and achieve a smooth transition to the kinetic regime.

\section{Gradient expansion}
\label{sec_grad}
 
 We determine second order transport coefficients by matching the 
conserved currents in hydrodynamics to the currents in kinetic theory. 
In hydrodynamics the particle current is $\vec\jmath=n\vec{u}$. This 
relation defines the fluid velocity $\vec{u}$, and does not receive 
dissipative corrections. The stress tensor is 
\be 
 \Pi_{ij} = \rho u_i u_j + P\delta_{ij}+ \delta \Pi_{ij}\, ,
\ee
where $P$ is the pressure and $\delta\Pi_{ij}$ is the dissipative part
of the stress tensor. At first order  in the gradient expansion $\delta
\Pi_{ij}$ can be written as $\delta\Pi_{ij}=-\eta\sigma_{ij}-\zeta\delta_{ij}
\langle \sigma\rangle$ with
\be 
 \sigma_{ij} = \nabla_i u_j +\nabla_j u_i 
  -\frac{2}{3}\delta_{ij}   \langle\sigma\rangle \, ,
\hspace{0.1\hsize}
 \langle\sigma\rangle =\vec{\nabla}\cdot\vec{u}\, ,
\ee
where $\eta$ is the shear viscosity and $\zeta$ is the bulk viscosity.
In a scale invariant fluid $\zeta=0$ \cite{Son:2005tj}. The general 
structure of dissipative corrections in a scale invariant fluid up to 
second order in the gradient expansion was studied in \cite{Chao:2011cy}. 
The result is 
\bea 
\delta\Pi_{ij} &=& -\eta\sigma_{ij}
   + \eta\tau_R\left[
      \dot\sigma_{ij} + u^k\nabla_k \sigma_{ij}
    + \frac{2}{3} \langle \sigma\rangle \sigma_{ij} \right] 
    + \lambda_1 \sigma_{\langle i}^{\;\;\; k}\sigma^{}_{j\rangle k} 
    + \lambda_2 \sigma_{\langle i}^{\;\;\; k}\Omega^{}_{j\rangle k}\nonumber\\
   && \mbox{} 
    + \lambda_3 \Omega_{\langle i}^{\;\;\; k}\Omega^{}_{j\rangle k}  
    + \gamma_1 \nabla_{\langle i}T\nabla_{j\rangle}T
    + \gamma_2 \nabla_{\langle i}P\nabla_{j\rangle}P
    + \gamma_3 \nabla_{\langle i}T\nabla_{j\rangle}P  \nonumber \\[0.1cm]
   \label{del_pi_fin}
   && \mbox{}
    + \gamma_4 \nabla_{\langle i}\nabla_{j\rangle}T 
    + \gamma_5 \nabla_{\langle i}\nabla_{j\rangle}P \, . 
\eea
Here, ${\cal O}_{\langle ij\rangle}=\frac{1}{2}({\cal O}_{ij}+{\cal O}_{ji}
-\frac{2}{3}\delta_{ij}{\cal O}^k_{\;\;k})$ denotes the symmetric traceless 
part of a tensor ${\cal O}_{ij}$, and $\Omega_{ij} = \nabla_iu_j-\nabla_ju_i$ 
is the vorticity tensor. In this work we focus terms related to gradients
of the velocity field and determine $\tau_R$ and $\lambda_i$. We will 
discuss the physical significance of these parameters in Sec.~\ref{sec_out}.

\section{Kinetic theory and the Chapman-Enskog method}
\label{sec_ce}

 In kinetic theory the conserved currents are expressed in terms
of quasi-particle distribution functions $f_p(\vec{x},t)$. The mass
current is 
\be 
\vec{\jmath} = \int d\Gamma_p \, m\vec{v}f_p(\vec{x},t)\, , 
\ee
where $d\Gamma_p=(d^3p)/(2\pi)^3$, $m$ is the mass of the particles, 
$\vec{v}=\vec{\nabla}_pE_p$ is the quasi-particle velocity, and $E_p$ 
is the quasi-particle energy. We will compute transport coefficients at 
leading order in the fugacity $z=\exp(\mu/T)$. For this purpose we can 
approximate $E_p$ by the energy of a free fermion, $E_p=p^2/(2m)$. The 
stress tensor is given 
by 
\be
 \Pi_{ij} = \int d\Gamma_p\, v_i p_j f_p(\vec{x},t)\, .
\ee
The distribution function is determined by the Boltzmann equation
\be
\label{be}
\left( \partial_t + \vec{v}\cdot\vec{\nabla}_x 
                  - \vec{F}\cdot\vec{\nabla}_p \right) 
  f_p(\vec{x},t) = C[f_p]\, , 
\ee
where $\vec{F}=-\vec{\nabla}_xE_p$. In the dilute limit $E_p$ is not
a function of $\vec{x}$ and we can set $\vec{F}=0$. We can write 
the Boltzmann equation as ${\cal D}f_p=C[f_p]$, where we have defined
\be 
 {\cal D} = \partial_t + \vec{v}\cdot\vec{\nabla}_x\, . 
\ee
At leading order in the fugacity the collision term is dominated by 
two-body collisions and the effects of quantum statistics can be 
neglected. We have 
\be 
 C[f_1]= -\prod_{i=2,3,4}\Big(\int d\Gamma_{i}\Big) w(1,2;3,4)
   \left( f_1f_2-f_3f_4\right)\, , 
\ee
where $f_i=f_{p_i}$ and $w(1,2;3,4)$ is the transition probability 
for $\vec{p}_1+\vec{p}_2\to\vec{p}_3+\vec{p}_4$. In the dilute Fermi
gas the scattering amplitude is dominated by s-wave scattering and
\be
w(1,2;3,4) = (2\pi)^4\Big(\sum_i E_i\Big)
         \delta\Big(\sum_i \vec{p}_i\Big) \,|{\cal A}|^2\, ,  
 \hspace{0.5cm} 
      |{\cal A}|^2 = \frac{16\pi^2}{m^2}\frac{a^2}{q^2a^2+1}\, , 
\ee
where $a$ is the $s$-wave scattering length and $2\vec{q}=\vec{p}_2
-\vec{p}_1$. The collision term vanishes in local thermal equilibrium, 
corresponding to the distribution function
\be 
\label{f_MB}
 f^{0}_p= \exp\Big(\frac{\mu}{T}\Big)
      \exp\Big(-\frac{m\vec{c}^{\,2}}{2T}\Big)\, , 
\ee
where $\vec{c}=\vec{v}-\vec{u}$, and the thermodynamic variables $\mu,T$ 
and $\vec{u}$ are functions of $\vec{x}$ and $t$. We will solve the 
Boltzmann by expanding the distribution function $f_p$ around the local 
equilibrium distribution, 
\be 
 f_p= f_p^{0} + f_p^{1} + f_p^{2} + \ldots 
    =  f_p^{0}\Big( 1 + \frac{\psi^{1}_p}{T}  + \frac{\psi^{2}_p}{T} 
     + \ldots \Big) \, . 
\ee
Inserting this ansatz into the Boltzmann equation gives
\be 
\label{ce}
 {\cal D}f_p^{0} + {\cal D}f_p^{1} + \ldots  
 = \frac{f_p^{0}}{T} \Big( 
      C^{1}_L\left[\psi^{1}_p\right] +  C^{2}_L\left[\psi^{1}_p\right]
    + C^{1}_L\left[\psi^{2}_p\right] + \ldots \Big)\, , 
\ee
where we have linearized the collision term, 
\bea
  C^{1}_L\left[\psi_1\right] &=&  
  -\int\Big(\prod_{i=2,3,4}d\Gamma_{i}\Big) w(1,2;3,4)\, f^0_2\,
      \left( \psi_1+\psi_2-\psi_3-\psi_4 \right)\, ,  \\  
  C^{2}_L\left[\psi_1\right] &=&
  -\int\Big(\prod_{i=2,3,4}d\Gamma_{i}\Big) w(1,2;3,4)\, \frac{f^0_2}{T}\,
      \left( \psi_1\psi_2-\psi_3\psi_4 \right)\, .
\eea
The left hand side of equ.~(\ref{ce}) represents an expansion in gradients 
of the thermodynamic variables, and the right hand side is an expansion in 
powers of the inverse mean free path $l_{\it mfp}^{-1}$. The dimensionless 
expansion parameter is the Knudsen number ${\it Kn}=l_{\it mfp}/L$, where 
$L$ is the characteristic length scale associated with the spatial 
dependence of the thermodynamic variables. 

 We will solve the Boltzmann equation to second order in the Knudsen 
number. At first order we have 
\be 
\label{psi_1}
 \psi^{1}_p = \left(C_L^1\right)^{-1} X^0_p\, ,\hspace{0.4cm}
 X^0_p\equiv \frac{T}{f^0_p}\left({\cal D}f^0_p\right)\, . 
\ee
Given the function $\psi^1_p$ the dissipative contribution to the stress 
tensor at first order in the gradient expansion is determined by 
\be 
 \delta \Pi^1_{ij} = \frac{\nu m}{T}\langle v_iv_j | \psi^1_p\rangle \, , 
\ee
where $\nu=2$ is the spin degeneracy and we have used $p_i=mv_i$. We 
have defined the inner product 
\be
\label{braket}
 \langle \psi_p|\chi_p\rangle = \int d\Gamma_p \, f^0_p \,
          \psi_p\chi_p \, .
\ee
The symmetries of the collision term imply that $C_L^1$ is a hermitean 
operator with respect to this inner product. In a scale invariant fluid 
$\delta \Pi_{ij}$ is traceless and we can replace $v_{ij}\equiv v_iv_j$ 
by $\bar{v}_{ij}\equiv v_{ij}-\frac{\delta_{ij}}{3}v^2$. If scale invariance 
is broken then $\langle \bar{v}_{ij} | \psi^1_p\rangle$ projects out the 
traceless part of the stress tensor. 

At next order in the Knudsen expansion the solution of the 
Boltzmann equation is 
\be 
\label{psi_2}
\psi^2_p =  \left(C_L^1\right)^{-1} \Big\{
   \left( X^0_p-C^1_L\left[\psi^1_p\right]\right)
 + X^1_p - C_L^2\left[\psi^1_p\right] \Big\}
\, ,\hspace{0.4cm}
 X^1_p\equiv \frac{T}{f^0_p}\left({\cal D}f^1_p\right)\, . 
\ee
The second order contribution to $\delta\Pi_{ij}$ is determined 
by the matrix element $\langle v_ip_j|\psi^2_p\rangle$. Because
$C_L^1$ is hermitean we can write 
\be 
\label{pi_ij_2}
 \delta \Pi^2_{ij} = \frac{\nu m}{T}
  \big\langle \left(C_L^1\right)^{-1}\bar{v}_{ij} \big|
  \Delta X^0_p + X^1_p - C_L^2\left[\psi^1_p\right] \big\rangle \, ,
\ee
where $\Delta X^0_p = X^0_p-C^1_L\left[\psi^1_p\right]$. Equ.~(\ref{psi_1})
implies that $\Delta X^0_p = 0$ at first order in the gradient expansion,
but in general $\Delta X^0_p$ is non-vanishing at $O(\nabla^2)$. The
advantage of letting $(C_L^1)^{-1}$ act to the left is that $C_L$ is 
an integral transformation which is not easy to invert. Indeed, solving
equ.~(\ref{psi_2}) using the exact collision operator is quite 
involved. However, computing $(C_L^1)^{-1}\bar{v}_{ij}$  is essentially
equivalent to computing $(C_L^1)^{-1}X_p^0$ and requires no extra
effort once $\psi^1_p$ is determined. 

 The calculation of $\delta \Pi^2_{ij}$ involves the following steps: 
1) Solve the first order equation for $\psi^1_p$. The solution to this
problem is well known \cite{Bruun:2005}.
2) Compute the second order streaming terms $ \Delta X^0_p$ and
$X^1_p$. Some of the required calculations can be found \cite{Chao:2011cy}.
3) Determine the second order collision term $C_L^2\left[\psi^1_p\right]$.
4) Compute the matrix element in equ.~(\ref{pi_ij_2}). 
We will go through these steps in the following sections. 

\section{First order solution}
\label{sec_first}

 We begin by computing the first order streaming term $X^0_p= \frac{T}{f^0_p}
({\cal D}f^0_p)$. Using equ.~(\ref{f_MB}) we find
\bea
\label{X^0}
X^0_p & =& \frac{m}{2}\, \bigg\{ 
   2\, \frac{T}{m}\, {\cal D}_u \alpha 
 + 2c^i \left[ {\cal D}_u u_i + \frac{T}{m} \nabla_i\alpha \right]  \\
 & & \hspace{0.7cm}\mbox{} 
 + c^ic^j\left[ \sigma_{ij} + \delta_{ij} \left( {\cal D}_u \log(T) 
      + \frac{2}{3}\langle \sigma\rangle \right)\right]
 + c^2c^k\nabla_k \log(T) \bigg\}\, , \nonumber 
\eea
where ${\cal D}_u=\partial_0+\vec{u}\cdot\vec{\nabla}$ is the comoving 
derivative relative to the fluid velocity $\vec{u}$, and we have defined 
$\alpha=\mu/T$. This equation can be simplified using the equations of 
fluid dynamics. Since equ.~(\ref{X^0}) is first order in gradients we 
can neglect gradient terms in the hydrodynamic equations. To leading order 
in the fugacity we can also use the equation of state of a free gas, 
$P=nT$. The continuity equation, the Euler equation, and the equation
of energy conservation are
\be 
\label{euler}
 {\cal D}_u\alpha=0\, , \hspace{0.6cm}
 {\cal D}_uu_i + \frac{T}{m} \nabla_i\alpha = 
   -\frac{5T}{2m}\nabla_i\log(T)\, , \hspace{0.6cm}
 {\cal D}_u\log(T) + \frac{2}{3}\langle \sigma\rangle=0\, ,
\ee
which leads to 
\be
\label{X^0_eq2}
X^0_p = \frac{m}{2}\, \bigg\{ 
   c^ic^j \sigma_{ij} 
  + c^k \left[\frac{5T}{m}- c^2 \right] q_k \bigg\}\, , 
\ee
where we have defined $q_k=-\nabla_k\log(T)$. Note that equ.~(\ref{X^0_eq2})
is orthogonal to the zero modes of the collision operator, $\langle X^0_p 
|\phi^{0,k}\rangle$ with $\phi^{0,k} =\{1,\vec{c},c^2\}$ $(k=1,2,3)$. 
These zero modes are associated with conservation of particle number, 
momentum, and energy. 

 As explained in the introduction we will focus on the case of no heat 
flow, $q_k=0$. In order to solve the Boltzmann equation $|X^0_p\rangle 
=C_L^1 |\psi^1_p\rangle$ we make an ansatz for $\psi^1_p$, 
\be 
 \psi^1_p = \sum_k^{N-1} a_k S_k\left(x_c\right)
   \bar{c}^{ij}\sigma_{ij}\, , 
   \hspace{0.5cm}
    x_c = \frac{mc^2}{2T}\, , 
\ee
where $S_k(x)$ is a complete set of functions. In practice we choose
$S_k(x)=L_k^{5/2}(x)$, where $L_k^{5/2}$ is a generalized Laguerre 
(Sonine) polynomial. This choice is convenient because of the orthogonality 
relation $\langle S_k\, \bar{c}^{ij}|S_l\, \bar{c}_{ij}\rangle
\sim \delta_{kl}$. We determine the coefficients $a_k$ by computing 
moments of the Boltzmann equation
\be 
\label{BE_mom}
 \big\langle S_k\, \bar{c}^{ij} \big| (X^0_p)_{ij} \big\rangle 
  =  \big\langle S_k\, \bar{c}^{ij} \big| C_L^1 
         \big| (\psi^1_p)_{ij}\big\rangle \, ,
  \hspace{0.5cm}(k=0,\ldots, N-1) \, ,
\ee
where we have defined $X^0_p=(X^0_p)^{ij}\sigma_{ij}$ and $\psi^1_p=
(\psi^1_p)^{ij}\sigma_{ij}$. From equ.~(\ref{X^0_eq2}) we get $(X^0_p)^{ij}
=\frac{m}{2}\,\bar{c}^{ij}$. As a first approximation we can set $N=1$, 
so that $\psi^1_p=a_0\,\bar{c}^{ij}\sigma_{ij}$ with 
\be 
a_0 =\frac{m}{2}\,\frac{\big\langle \bar{c}^{kl} \big|\bar{c}_{kl}\big\rangle}
 {\big\langle \bar{c}^{ij} \big| C_L^1  \big| \bar{c}_{ij}\big\rangle}\, . 
\ee
The matrix element of the collision operator is 
\be 
\big\langle \bar{c}_{ij} \big| C_L^1  \big| \bar{c}^{ij}\big\rangle
 =  -\int\Big(\prod_{i=1}^4 d\Gamma_{i}\Big) w(1,2;3,4)\, f^0_1 f^0_2\,
     \left(\bar{c}_1\right)_{ij}\, 
         \left( \bar{c}_1^{ij}+\bar{c}_2^{ij}-\bar{c}_3^{ij} - \bar{c}_4^{ij}  
         \right)\, . 
\ee
At unitarity this integral can be computed analytically, see 
Sec.~\ref{sec_coll}. We find
\be 
 a_0 \equiv \bar{a}_0 \, \frac{m}{zT}
     = -\frac{15\pi}{32\sqrt{2}} \, \frac{m}{zT}\, ,
\ee
and the shear viscosity is 
\be
\label{eta_kin}
 \eta = \frac{15}{32\sqrt{\pi}}\, (mT)^{3/2}\, . 
\ee
Using $(C_L^1)^{-1}|(X^0_p)_{ij}\rangle = |(\psi^1_p)_{ij}\rangle$
together with $(X^0_p)_{ij}=\frac{m}{2}\,\bar{c}_{ij}$ we observe that 
this result determines the quantity $(C_L^1)^{-1}|\bar{v}_{ij}\rangle$
which enters in equ.~(\ref{pi_ij_2}). We find
\be 
\label{C_L_inv}
\left(C_L^1\right)^{-1}\big|\bar{v}_{ij}\big\rangle = 
\big|\bar{v}_{ij}\big\rangle  \frac{2}{zT}\,\bar{a}_0  \, , 
\ee
which is correct up to higher order terms in the Sonine polynomial 
expansion. The next-to-leading order correction is determined in App.~A.

\section{Second order streaming terms}
\label{sec_second}

 Once $f^1_p=f^0_p\psi^1_p/T$ is determined we can compute the second
order streaming term $X^1_p=(T/f^0_P)({\cal D}f^1_p)$. The Boltzmann
equation implies that the sum $X^1_p+\Delta X^0_p$ must be orthogonal 
to the zero modes of the collision operator, but the two terms do 
not satisfy the orthogonality constraints individually. We can decompose 
$X^1_p=(X^1_p)_{\it orth}+(X^1_p)_{\it ct}$, where $(X^1_p)_{\it orth}$ is 
orthogonal to the zero modes, and $(X^1_p)_{\it ct}$ is a ``counterterm''
that will have to cancel against contributions contained in $\Delta X^0_P$. 
We find
\bea
\label{X^1_orth}
\left(X^1_p\right)_{\it orth}  &=&  \frac{m\bar{a}_0}{zT} \left\{
 \frac{m}{2T} \Big( c^ic^jc^kc^l 
     -\frac{2}{15} \delta^{ik}\delta^{jl}c^4\Big)
           \sigma_{ij}\sigma_{kl}\right.
  \\
 & & \hspace{0.6cm}\mbox{} 
    +\Big( c^ic^j-\frac{1}{3}\delta^{ij}c^2\Big) 
     \left[ \Big({\cal D}_u+\frac{2}{3}\langle\sigma\rangle\Big)\sigma_{ij}
         - \sigma^{}_{ik}\sigma_{j}^{\;\;k} 
         - \sigma^{}_{ik}\Omega_{j}^{\;\;k} \right]
   \nonumber \\
& & \hspace{0.6cm}\mbox{} \left.
   +\Big( c^ic^jc^k-\frac{3}{5}\delta^{(ij}c^{k)}c^2\Big)
    \, \nabla_k\sigma_{ij} 
    \right\}\, , \nonumber
\eea
where $\Omega_{ij}=\nabla_iu_j-\nabla_ju_i$ is the vorticity tensor, 
and $A_{(ijk)}$ is symmetrized in all tensor indices. We have dropped
two-derivative terms proportional to gradients of $T$ and $\alpha$. 
The counterterms are 
\be
\left(X^1_p\right)_{\it ct}  =  \frac{m\bar{a}_0}{zT} \left\{
      \Big( \frac{m}{15T}\, c^4 - \frac{1}{3}\, c^2 \Big) \sigma^2
   + \frac{3}{5}\delta^{(ij}c^{k)}c^2
    \, \nabla_k\sigma_{ij}  \right\}\, , 
\ee
where $\sigma^2=\sigma^{ij}\sigma_{ij}$. The second streaming term, 
$\Delta X^0_p$, can be determined as in Sect.~\ref{sec_first}, but 
at second order in the gradient expansion we have to use the 
Navier-Stokes equation rather than the Euler equation. We have
\bea 
\label{ns_1}
 {\cal D}_u\alpha\;  &=& -\frac{\eta}{2P}\, \sigma^2\, , \\
\label{ns_2}
 {\cal D}_u u_i   \; &=& - \frac{T}{m} \left( \nabla_i\alpha  
   +\frac{5}{2} \nabla_i\log(T) \right)
   -\frac{1}{\rho}\, \nabla^k \left(\eta \sigma_{ki}\right) \, , \\
\label{ns_3}
 {\cal D}_u\log(T)   &=&  - \frac{2}{3}\langle \sigma\rangle 
   +\frac{\eta}{3P}\, \sigma^2 \, ,
\eea
where we have neglected second order terms that involve gradients
of the temperature. In order to be consistent with the first order
calculation we use the result for $\eta$ found in Sect.~\ref{sec_first}. 
We can write 
\be
\label{eta_a_0}
 \eta = -\frac{\sqrt{2}}{\pi^{3/2}}\, \bar{a}_0 (mT)^{3/2}\, ,
\ee
as well as $P=nT$ and $\rho=mn$. Here, $n=\nu z\lambda^{-3}$ is the 
density and $\lambda=[(2\pi)/(mT)]^{1/2}$ is the thermal de Broglie wave 
length. Combining equ.~(\ref{X^0}) with the Navier-Stokes equation
(\ref{ns_1}-\ref{ns_3}) we find
\be 
\Delta X^0_p = -\frac{m\bar{a}_0}{zT} \left\{
      \left( \frac{1}{3}  c^2 - \frac{T}{m}  \right) \sigma^2
   + \frac{2T}{m} \, c^i\, \nabla_k\sigma_{ki}  \right\}\, .
\ee
We observe that $\Delta X^0_p$ is a sum of terms that are proportional 
to zero modes of the collision operator, but that it does not cancel 
against $(X^1_p)_{\it ct}$. In particular, $(X^1_p)_{\it ct}$ contains terms 
of order $c^4$ and $c^3$, whereas $\Delta X^1_p$ is a second order 
polynomial in $c$. We can, however, write any polynomial in $c$ as
the sum of a polynomial orthogonal to the zero modes, and a polynomial
of lower order in $c$. In particular, we can write
\begin{align}
\label{ortho1}
c^4    &= \chi_4 + \left( c^4-\chi_4 \right)\, , 
&\chi_4 &= c^4 - 10\, \frac{T}{m}\, c^2 + 
                    15 \left(\frac{T}{m}\right)^2\, , \\
\label{ortho2}
c_ic^2 &= \chi_{3,i} + \left(c_i c^2-\chi_{3,i} \right)\, ,
&\chi_{3,i} &= c_i \left( c^2 - 5\, \frac{T}{m}\right)\, , 
\end{align}
where $\langle \chi_4 | \phi^{0,k} \rangle = \langle\chi_{3,i} | \phi^{0,k} 
\rangle=0$ with $\phi^{0,k}=\{1,c_i,c^2\}$ $(k=1,2,3)$. Using 
equ.~(\ref{ortho1},\ref{ortho2}) we can write $(X_p^1)_{\it ct}= 
(X_p^1)_{\it ct,orth}+(X_p^1)_{\it ct,zm}$ where $(X_p^1)_{\it ct,orth}$ 
is orthogonal to the zero modes. We find $\Delta X^0_p+(X_p^1)_{\it ct,zm}=0$, 
and the second order streaming term is orthogonal to the zero modes. The 
complete streaming term at second order in the gradient expansion is
\be
X^1_p + \Delta X^0_p= (X^1_p)_{\it orth}+(X_p^1)_{\it ct,orth}\, , 
\ee
where $(X^1_p)_{\it orth}$ is given in equ.~(\ref{X^1_orth}) and 
\be 
\left(X^1_p\right)_{\it ct,orth}  =  \frac{m\bar{a}_0}{zT} \left\{
      \frac{m}{15T}\left( c^4 - \frac{10T}{m}\, c^2 
          +  \frac{15T^2}{m^2} \right) \sigma^2
   + \frac{2}{5} \left( c^2-\frac{5T}{m} \right) \,
       c^{i} \nabla_j\sigma_{ij}  \right\}\, .
\ee
In order to determine the transport coefficients we need to compute 
scalar products with $|c_{ij}\rangle$. Because of the orthogonality 
properties of $\chi_4$ and $\chi_{3,i}$ we find 
\be 
 \big\langle c_{ij} \big| \left(X^1_p\right)_{\it ct,orth} \big\rangle 
= \frac{1}{3} 
 \big\langle c^2 \big| \left(X^1_p\right)_{\it ct,orth} \big\rangle 
= 0\, ,
\ee
and only $(X^1_p)_{\it orth}$ contributes to $\delta\Pi^{2}_{ij}$.

\section{Second order collision term}
\label{sec_coll}

 The second order collision operator is 
\bea
  C^{2}_L\left[\psi^1_1\right] &=&
  -\int d\Gamma_{234}\, w(1,2;3,4)\, f^0_2\,
      \left( \psi^1_1\psi^1_2-\psi^1_3\psi^1_4 \right) \nonumber \\
  & =& - \frac{1}{T}\left(\frac{\bar{a}_0m}{zT}\right)^2\, 
      \sigma_{ij}\sigma_{kl}
    \int d\Gamma_{234}\, w(1,2;3,4)\, f^0_2\,
     \left(\bar{c}_1^{ij}\bar{c}_2^{kl} 
           - \bar{c}_3^{ij}\bar{c}_4^{kl}\right) \, ,
\eea
where $d\Gamma_{234}=d\Gamma_2\,d\Gamma_3\, d\Gamma_4$ and we have 
used $\psi^1_p=\bar{a}_0m(zT)^{-1}\bar{c}^{ij}\sigma_{ij}$. In order 
to determine the stress tensor we need
\be 
 \big \langle \bar{c}_1^{ab} \big|   
                 C^{2}_L\left[\psi^1_1\right]\big\rangle 
 \equiv  \left(\frac{\bar{a}_0m}{zT}\right)^2\, 
       (C^2_L)^{abijkl} \sigma_{ij}\sigma_{kl} \,\, . 
\ee 
$(C^2_L)_{abijkl}$ is a rank 6 tensor which is symmetric and traceless
in $(ab)$, $(ij)$ and $(kl)$, and symmetric under the exchange $(ij)
\leftrightarrow (kl)$. These symmetries completely fix the tensor structure
and we find
\be 
(C^2_L)_{abijkl}\, \sigma^{ij}\sigma^{kl} 
    = \frac{12}{35}\, {\cal C}^2_L \,
      \sigma_{\langle a}^{\;\;\; c}\sigma^{}_{b\rangle c} \, ,
\ee
where we have defined the scalar integral
\be
{\cal C}^2_L \equiv (C^2_L)_{ab\;\;c}^{\;\;\;\; a\;\; cb} 
 = -\int d\Gamma_{1234}\, w(1,2;3,4)\, f^0_1f^0_2\,
     \left(\bar{c}_1\right)_{ab}
     \left[ \left(\bar{c}_1\right)^{a}_{\;\,c}
            \left(\bar{c}_2\right)^{cb} 
          - \left(\bar{c}_3\right)^{a}_{\;\,c}
            \left(\bar{c}_4\right)^{cb}     \right]\, . 
\ee
The scalar collision integral can be computed in analogy with the first 
order collision integral. We introduce center-of-mass and relative momenta
\be 
 m\,\vec{c}_{1,2} = \frac{\vec{P}}{2}\pm \vec{q} \, , \hspace{0.5cm}
 m\,\vec{c}_{3,4} = \frac{\vec{P}}{2}\pm \vec{q}^{\,\prime}\, , 
\ee
and write the phase space measure as
\begin{align} 
\int d\Gamma_{1234} \, (2\pi)^4 & \delta^3\Big(\sum_i \vec{p}_i\Big)
    \delta \Big(\sum_i E_i\Big)\nonumber \\
  & = \frac{2}{(2\pi)^6} \int P^2dP \int q^2 dq \,\frac{qm}{2}\, 
      \int d\cos\theta_q\int d\cos\theta_{q'}\int d\phi_{q'}\, , 
\end{align}
where we have chosen a coordinate system in which $\vec{P}=P\hat{z}$,
so that $\hat{P}\cdot\hat{q}=\cos\theta_{q}$. We also have $\hat{P}
\cdot\hat{q}'=\cos\theta_{q'}$ and $\hat{q}\cdot\hat{q}'=\cos\theta_{q}
\cos\theta_{q'}+\sin\theta_{q}\sin\theta_{q'}\cos\phi_{q'}$. Neither the 
product of distribution functions, $f_1^0f_2^0$, nor the scattering 
amplitude, $|{\cal A}|^2$, depend on the angles $\theta_q,\theta_{q'}$ 
and $\phi_{q'}$. The angular integral can be performed by symmetrizing the 
integrand
\begin{align} 
\int d\cos\theta_q & \int d\cos\theta_{q'}\int d\phi_{q'}
    \; \frac{1}{8}\Big[
      \left(\bar{c}_1\right)_{ab} + \left(\bar{c}_2\right)_{ab}
     -\left(\bar{c}_3\right)_{ab} - \left(\bar{c}_4\right)_{ab}
              \Big] \nonumber \\
  & \times \Big[ \left(\bar{c}_1\right)^{a}_{\;\,c}
             \left(\bar{c}_2\right)^{cb}
           - \left(\bar{c}_3\right)^{a}_{\;\,c}
             \left(\bar{c}_4\right)^{cb}     
           + (1\leftrightarrow 2,3\leftrightarrow 4)     
      \Big]\,
 = \frac{4\pi}{27}\, \frac{q^4}{m^6} \left(12q^2-7P^2\right)\, . 
\end{align}
The integrals over $P$ and $q$ factorize and
\begin{align}
{\cal C}^2_L &= -\frac{2}{(2\pi)^6}
  \int P^2dP\, \int q^2dq\,  f_1^0f_2^0\, \frac{mq}{2} \,
  \left[ \frac{4\pi}{27}\, \frac{q^4}{m^6} \left(12q^2-7P^2\right)\right]
   \, \frac{16\pi^2}{m^2q^2}  \nonumber \\
  & =   \frac{4T^{11/2}}{9\pi^{5/2}m^{3/2}}\, , 
\end{align}
where we have taken the unitary limit, $a\to\infty$. This result 
determines the matrix element of the second order collision term, 
\be
\label{c_ij_C^2}
 \big \langle \left(\bar{c}_1\right)_{ij} \big|   
                 C^{2}_L\left[\psi^1_1\right]\big\rangle 
    = \frac{16\,\bar{a}_0^2\, m^{1/2}T^{5/2}}{105\pi^{5/2}}\, 
        \sigma_{\langle i}^{\;\;\; k}\sigma^{}_{j\rangle k}\, . 
\ee

\section{Stress tensor and second order transport coefficients}
\label{sec_stress}

 We now have all the ingredients in place to compute the dissipative
stress tensor at second order in the gradient expansion. We start from
equ.~(\ref{pi_ij_2}) and use equ.~(\ref{C_L_inv}) to compute $(C_L^1)^{-1}
\bar{c}_{ij}$. We find
\be 
\label{pi_ij_2_b}
 \delta \Pi^2_{ij} = \frac{2\nu m\bar{a}_0}{zT^2}
  \Big(
    \big\langle \bar{c}_{ij} \big| (X^1_p)_{\it orth}\big\rangle
  - \big\langle \bar{c}_{ij} \big| C_L^2\left[\psi^1_p\right] \big\rangle 
   \Big)\, .
\ee
The projection of the collision term on $|\bar{c}_{ij}\rangle$ is given 
in equ.~(\ref{c_ij_C^2}). The projection of the streaming term is 
\be 
 \big\langle \bar{c}_{ij} \big|
         (X^1_p)_{\it orth} \rangle  = 
  \frac{\bar{a}_0 m^{1/2}T^{5/2}}{\sqrt{2}\pi^{3/2}} \left[
      \Big( {\cal D}_u + \frac{2}{3}\langle\sigma\rangle\Big) \sigma_{ij}
     + \sigma_{\langle i}^{\;\;\; k}\sigma^{}_{j\rangle k}
     - \sigma_{\langle i}^{\;\;\; k}\Omega^{}_{j\rangle k} \right]\, . 
\ee
We can read off the transport coefficients by matching 
equ.~(\ref{pi_ij_2_b}) to the general result in conformal fluid 
dynamics, equ.~(\ref{del_pi_fin}). We use equ.~(\ref{eta_a_0})
to relate $\bar{a}_0$ to the shear viscosity. We find
\be 
\label{final}
\tau_R=\frac{\eta}{P}\, , \hspace{0.3cm}
\lambda_1 =  \frac{15\eta^2}{14P}\, , \hspace{0.3cm}
\lambda_2 = -\frac{\eta^2}{P}\, , \hspace{0.3cm}
\lambda_3 = 0 \, . 
\ee
This result is exact at leading order in the fugacity and Sonine 
polynomial expansion. It is straightforward to compute higher order 
terms in the Sonine polynomial expansion. For the shear viscosity 
this is done in App.~\ref{sec_nlo}. This solution can be inserted 
into the second order streaming and collision terms in order to 
compute the next order correction to $\tau_R$ and $\lambda_i$. 
We can see that the relation $\tau_R=\eta/P$ is not modified. The 
coefficients $\lambda_i$ receive corrections that are parametrically 
of the same magnitude as the corrections to $\eta$, which is less
than 2\%. Higher order corrections in the fugacity expansion are more 
difficult to compute. These corrections include higher order corrections 
to the equation of state and the quasi-particle properties, quantum
effects, and three body collisions. Estimates of these effects can be 
obtained from the $T$-matrix calculation described in \cite{Enss:2010qh} 
and the molecular dynamics simulation in \cite{Dusling:2012fi}. Both
calculations show that corrections to the dilute limit become large
for $T\lsim T_F$, and that these effects tend to increase the shear
viscosity. 

 Equ.~(\ref{final}) can be compared to the result of the relaxation
time (BGK) approximation \cite{Chao:2011cy}. In this case we replace
the full collision operator by $C[f^0+\delta f]\simeq -\delta f/\tau_0$.
This is a very crude approximation, but one that has been successfully
applied in many areas of kinetic theory. Of course, one can always 
choose $\tau_0$ to obtain the correct shear viscosity, but the error 
in other transport properties is not necessarily small. We find, however, 
that $\tau_R$ and $\lambda_{2,3}$ agree with the BGK approximation, and 
that the correction in $\lambda_1$, the factor 15/14 in equ.~(\ref{final}),
is close to one. 

 The reason that $\lambda_1$ is modified is easily traced to the fact
that $\psi^1_p\sim c_{ij}\sigma^{ij}$, so that the non-linear collision
term generates terms proportional to $\sigma_{i}^{\;\;k}\sigma_{jk}^{}$.
The fact that numerically this correction is small is essentially an
accident, which depends on the structure of the collision cross section.
The result that $\tau_R$ and $\lambda_2$ are not modified is somewhat 
harder to understand. The two main reasons are that the structure of 
$\psi^1_p$ is correctly reproduced by the BGK approximation, and that 
the second order streaming term ${\cal D}f^1_p$ is constrained by scale
invariance. Indeed, the BGK approximation leads to the correct relaxation
time $\tau_R$ in units of $\eta/P$ provided the collision time scales as 
$\tau_0\sim T^{-1}h(\alpha)$ for any function $h$. 

 In this work we have not studied higher order corrections to heat flow. 
In this case the BGK approximation is expected to be less useful. If the 
collision time $\tau_0$ is fixed using the shear viscosity, then the thermal 
conductivity is too small by a factor 2/3 \cite{Chao:2011cy,Braby:2010ec}.
In addition to that, non-linearities in the collision term will give 
corrections to $q_iq_j$ terms in the stress tensor. 

\section{Discussion}
\label{sec_out}

 The main result of our study is equ.~(\ref{final}), which provides
the transport coefficients related to terms of order $O(\nabla^2 u)$
in the stress tensor of a unitary Fermi gas. The results are exact at 
leading order in the fugacity $z$. In order to study the physical 
significance of second order terms we note that it is possible to 
rewrite the equations of fluid dynamics as the Navier-Stokes equation
coupled to a relaxation equation for the dissipative stresses $\pi_{ij}
\equiv\delta\Pi_{ij}$. For this purpose we use the first order relation 
$\pi_{ij}=-\eta\sigma_{ij}$ and write equ.~(\ref{del_pi_fin}) as 
\cite{Chao:2011cy}
\be
\label{pi_rel}
\pi_{ij} = -\eta\sigma_{ij}
   - \tau_R\left[
     \dot\pi_{ij} + u^k\nabla_k \pi_{ij}
    + \frac{5}{3} \langle \sigma\rangle \pi_{ij} \right] 
    + \frac{\lambda_1}{\eta^2} \pi_{\langle i}^{\;\;\; k}\pi^{}_{j\rangle k} 
    - \frac{\lambda_2}{\eta} \pi_{\langle i}^{\;\;\; k}\Omega^{}_{j\rangle k}
    + \lambda_3 \Omega_{\langle i}^{\;\;\; k}\Omega^{}_{j\rangle k}  \, , 
\ee
where we have dropped terms of order $O(\nabla^2 T)$. Equation (\ref{pi_rel}) 
is easiest to solve in systems in which the time dependence is harmonic, and 
non-linear terms in the velocity are small, $(\nabla u)^2\ll\nabla\dot{u}$. In 
this case the relaxation time equation is solved by $\pi_{ij}=-\eta(\omega)
\sigma_{ij}$, where $\eta(\omega)=\eta/(1-i\omega\tau_R)$ is an effective, 
frequency dependent, viscosity. 

 The two conditions stated above are satisfied in the case of collective 
modes of a trapped Fermi gas \cite{Bruun:2007,Schaefer:2009px}. We consider 
the damping of the transverse breathing mode \cite{Kinast:2004b}. In order 
to study the sensitivity of the damping rate to the values of the transport 
coefficients we write $\eta=c_\eta (mT)^{3/2}$ and $\tau_R=c_\tau\eta/P$. In 
kinetic theory $c_\eta=15/(32\sqrt{\pi})$ and $c_\tau=1$, see 
equ.~(\ref{eta_kin}) and (\ref{final}). The damping rate is determined 
by the spatial integral over the frequency dependent shear viscosity.
We find \cite{Schaefer:2009px}
\be 
\label{Gamma_br}
\Gamma = - \frac{c_\eta\omega_\perp}{(3\lambda N)^{1/3}}
    \left(\frac{E_F}{E_0}\right)
    \left(\frac{T}{\bar{\omega}}\right)^3
    {\rm Li}_{-3/2} \left(
      -\frac{3N^2\bar{\omega}^2}{80c_\eta^2c_\tau^2\pi^3 T^4}\right)\, ,
\ee
where $\omega_\perp$ is the transverse trap frequency, $\bar{\omega}$ 
is the geometric mean of the trapping frequencies, and $\lambda =\omega_z
/\omega_\perp$. ${\it Li}$ is a polylogarithm, $N$ is the total number 
of particles and $E_0/E_F$ is the total energy per particle in units of 
the Fermi energy. At low temperature the damping rate scales as $\Gamma
\sim c_\eta T^3 \log(c_\eta c_\tau T^2)$. In this regime fluid dynamics 
is valid over most of the cloud, and there is only a weak, logarithmic, 
dependence on the second order coefficient $c_\tau$. In the high temperature 
limit we find $\Gamma \sim 1/(c_\eta c_\tau^2 T)$. In this case the 
dependence on the second order coefficient $c_\tau$ is more important 
than the dependence on $c_\eta$ and the gradient expansion is not 
valid. However, the result agrees with the prediction of the Boltzmann 
equation for a trapped gas \cite{Bruun:2007}. This implies that 
equ.~(\ref{Gamma_br}) smoothly interpolates between second order 
fluid dynamics and kinetic theory. In particular, we can view the
result $\eta(\omega)=\eta/(1-i\omega\tau_R)\simeq \eta + i\omega\eta
\tau_R$ as a resummation of the second order term that builds in 
the correct extrapolation to the limit $\tau_R\to\infty$. 

  Equation (\ref{Gamma_br}) was compared to data in 
\cite{Bruun:2007,Schaefer:2009px}, and it was found that the agreement
with experiment in the regime $0.3\lsim T/T_F \lsim 1$ is quite good. 
In the original studies equ.~(\ref{Gamma_br}) was derived using 
the BGK model, which is not a systematic approximation. What we have 
shown in the present work is that the same result can be derived 
from a reliable calculation based on kinetic theory and the fugacity
expansion.   

 The role of $\lambda_1$ and $\lambda_2$ can be studied by considering
the hydrodynamic expansion of a Fermi gas after release from a harmonic 
trap. The initial state is in hydrostatic equilibrium in an axisymmetric 
harmonic potential with $\omega_z\ll \omega_z$. Hydrodynamic expansion 
converts the asymmetry of the potential into differential acceleration 
and leads to transverse flow. Shear viscosity counteracts this effect 
and suppresses transverse expansion. We can obtain a qualitative 
understanding of the effects of dissipative terms by computing the 
stress tensor for the velocity field that solves the Euler equation 
for an expanding gas cloud \cite{Menotti:2002,Schaefer:2009px}. The 
velocity field is of the form $u_i(x,t)=\alpha_i(t)x_i$ (no sum over $i$), 
which is analogous to Hubble expansion in cosmology and to Bjorken expansion 
in relativistic heavy ion physics. In the case of a strongly deformed 
trap $\alpha\equiv \alpha_\perp \gg \alpha_z$. The solution to the Euler 
equation can be written as $\alpha(t)=\dot{b}(t)/b(t)$, where $b(t)$ is 
the transverse scale factor of the expansion. At early time $b(t)\simeq 
1 + \frac{1}{2}\omega_\perp^2 t^2$, and at late time $b(t)\simeq 
\sqrt{\frac{3}{2}}\omega_\perp t$. The strain tensor $\sigma_{ij}$ 
is diagonal, $\sigma_{ij}=\frac{2}{3} {\rm diag}(\alpha,\alpha,-2\alpha)$. 

\begin{enumerate}
\item At first order in the gradient expansion we compare the dissipative 
stresses $\delta\Pi_{ij}=-\eta\sigma_{ij}=-\frac{2\eta}{3} {\rm diag}
(\alpha,\alpha,-2\alpha)$ to the ideal stresses $\Pi_{ij}=P\delta_{ij}
+\rho u_iu_j$. We observe, as expected, that dissipative effects tend to 
suppress transverse expansion and accelerate longitudinal expansion. 

\item The coefficient $\lambda_1$ determines non-linearities in 
the stress-strain relation. In the case of anisotropic expansion
we find $\delta\Pi_{ij}^{2}=\frac{4\lambda_1}{9}{\rm diag}(-\alpha^2,
-\alpha^2,2\alpha^2)$ and $\lambda_1>0$ implies that viscous stresses 
are increased by second order effects.

\item The transport coefficient $\lambda_2$ only plays a role in 
rotating systems. An example is the expansion from a rotating trap 
studied in \cite{Clancy:2007}. The initial state supports a velocity 
field of the form $\vec{u}=(\beta z,0,\beta x)$, which carries non-zero 
angular momentum but no vorticity. The first order stress tensor is 
$\delta\Pi^1_{ij}=-2\beta\eta(\delta_{xi}\delta_{zj}+\delta_{xj}\delta_{zi})$. 
The main effect of viscosity is to convert a fraction of the initial 
irrotational flow to rigid rotation, and generate non-zero vorticity 
\cite{Schaefer:2009px}. At second order in the gradient expansion 
vorticity couples to transverse expansion and the angular momentum
carried by the irrotational flow. This leads to two effects, an 
enhancement of transverse flow in-plane versus out of the rotation
plane, and a further enhancement of rigid rotation.

\end{enumerate}

\section{Final remarks}
\label{sec_final}

 In this paper we have computed second order transport coefficients 
for a dilute Fermi gas. Second order transport properties were first
considered by Burnett, who computed $\psi^2_p$ and $\delta\Pi_{ij}^2$ 
for Maxwell molecules, which are classical particles subject to a 
repulsive $1/r^5$ force \cite{Burnett:1935}. The calculation presented 
in this work is substantially simpler than Burnett's. Part of the 
simplification is due to a more compact notation. We also avoid 
explicitly calculating $\psi^2_p$, and we focus on a simpler 
interaction, albeit one that can be realized experimentally. Finally, 
exact scale invariance reduces the number of independent kinetic 
coefficients. 

 Second order kinetic coefficients have also been computed for a 
relativistic quark gluon plasma \cite{York:2008rr}. The general 
structure of the result is very similar to the non-relativistic case. 
In particular, in the case of a quark gluon plasma one finds $\tau_R
\simeq 3\eta/(2P)$, and $\lambda_1>0$, $\lambda_2<0$ as well as 
$\lambda_3=0$. All these results refer to the weak coupling, kinetic, 
limit. Second order transport coefficients of a relativistic scale 
invariant plasma have been computed in the strong coupling limit 
using the AdS/CFT correspondence \cite{Baier:2007ix}. In this case 
one finds $\tau_R=(1-\log(2)/2)\eta/P$ and, again, $\lambda_1>0$, 
$\lambda_2<0$ and $\lambda_3=0$.

 Assessing the full impact of $\tau_R$, $\lambda_1$ and $\lambda_2$ 
on the non-equilibrium evolution of expanding Fermi gas clouds
will require numerical simulation similar to those reported 
in \cite{Schafer:2010dv}. This work is in progress.
 
 Acknowledgments: This work was supported in parts by the US Department 
of Energy grant DE-FG02-03ER41260. 

\appendix
\section{The shear viscosity at next-to-leading order in the Sonine 
polynomial expansion}
\label{sec_nlo}

 It is straightforward, if somewhat tedious, to go beyond leading
order in the Sonine polynomial expansion. At next-to-leading 
order we write 
\be 
 \left(\psi^1_p\right)_{ij} = 
   \left( a_0 + a_1S_1\left(x_c\right) + \ldots  \right)\bar{c}_{ij}\, , 
\ee
where $S_1(x)=\frac{7}{2}-x$ and $x_c=mc^2/(2T)$. We define the matrix 
elements
\be 
 (C_L^1)_{IJ} = \big\langle S_I\bar{c}^{ij} \big| C_L^1 \big| 
                            S_J\bar{c}_{ij} \big\rangle\, , 
\ee
as well as the normalization constants
\be 
  {\cal N}_I = \frac{m}{2}  \big\langle S_I\bar{c}^{ij} \big|
                            S_I\bar{c}_{ij} \big\rangle \, . 
\ee
The expansion coefficients $a_i$ are determined by equ.~(\ref{BE_mom}).
If we truncate the expansion at $N=1$ we get 
\bea
 a^{(1)}_0 &=& {\cal N}_0 \, \frac{(C_L^1)_{11}}
         { (C_L^1)_{00} (C_L^1)_{11} - (C_L^1)_{01}^2 } \, , \\
 a^{(1)}_1 &=& {\cal N}_0 \, \frac{-(C_L^1)_{01}}
         { (C_L^1)_{00} (C_L^1)_{11} - (C_L^1)_{01}^2 } \, .
\eea
This should be compared to the $N=0$ solution $a^{(0)}_0= {\cal N}_0 /
(C_L^1)_{00}$. The matrix elements $(C_L^1)_{IJ}$ can be computed using 
the methods described in Sect.~\ref{sec_coll}. Because of the orthogonality
relation $\langle S_k\, \bar{c}_{ij}|S_l\, \bar{c}_{ij}\rangle\sim 
\delta_{kl}$ the shear viscosity is determined by $a_0^{(N)}$. For 
$N=1$ we obtain \cite{Bruun:2006}
\be 
 \eta^{(1)} = \eta^{(0)} \frac{(C_L^1)_{00}(C_L^1)_{11}}
         { (C_L^1)_{00} (C_L^1)_{11} - (C_L^1)_{01}^2}
  = \eta^{(0)} \, \frac{193}{190}\, , 
\ee
which is a 2\% correction. Note that the Sonine polynomial expansion
is variational. In particular, the shear viscosity computed from the 
exact solution of the Boltzmann equation is larger or equal to the 
$N$'th order approximant. Also note that the $N=1$ correction to 
the distribution function is somewhat larger than the correction 
to the shear viscosity. We find $a^{(1)}_1/a^{(1)}_0=-12/193\simeq
-0.06$. The sign of $a^{(1)}_1/a^{(1)}_0$ implies that particle 
are pushed out to slightly larger momenta compared to the $N=0$
approximation $\psi^1_p\sim c^{ij}\sigma_{ij}$.


\end{document}